\documentstyle[pra,aps,graphicx]{revtex}
\begin{document}
\draft
\twocolumn[\hsize\textwidth\columnwidth\hsize\csname 
@twocolumnfalse\endcsname
\title{Generation of spatial antibunching with free propagating twin beams} 
\author{D. P. Caetano$^{*}$ and  P.H. Souto Ribeiro}
\address{Instituto de F\'{\i}sica, Universidade Federal do Rio de 
Janeiro, Caixa Postal 68528, Rio de Janeiro, RJ 21945-970, Brazil}
\date{\today}
\maketitle
\begin{abstract}
We propose and implement a novel method to produce a spatial anti-bunched field with free propagating twin beams from spontaneous parametric down-conversion. The method consists in changing the spatial propagation by manipulating the transverse degrees of freedom through reflections of one of the twin beams. Our method use reflective elements eliminating losses from absorption by the objects inserted in the beams. 
\end{abstract}
\pacs{42.50.Ar, 42.25.Kb}
]
\section{Introduction}

In contrast with light sources where the emitted photons tend to propagate gathered together, special kinds of light sources can produce photons tending to propagate separated in time or space. This effect, referred to as photon antibunching, has been predicted in time\cite{1} and spatial\cite{1a} domain and observed in resonance fluorescence by excited atomic beams\cite{2}, trapped ions\cite{4}, trapped molecules\cite{5}, and in optical parametric process\cite{6,7}. Recently, Nogueira,  Walborn, P\'adua, and Monken\cite{7} have proposed and implemented a spatial version of the photon antibunching where photons tend to propagate separated with respect to the transverse propagation plane. They have used twin photons from the parametric down-conversion and
in their experiment, the final state is prepared by propagation through special double-slits, 
leading to a light beam in which photons seems to repeal each other in space.

In this work, we propose and implement an experimental scheme for obtaining spatial antibunching with free-propagating beams. It is based on the transfer of the angular spectrum from the pump to the twin beams of the down-conversion\cite{8} and the manipulation of the transverse spatial coordinate of one of the beams. We obtain a high quality anti-bunched beam. The preparation of the state is very efficient with respect to the incoming signal and idler beams.

\section{State preparation}

In order to produce spatial antibunching, one light beam must be able to violate a
classical inequality for its fourth order correlation function in the spatial
variables(See Eq. 11 of Ref.\cite{7}):

\begin{equation}
\label{eq1}
\Gamma^{(2,2)}(\bbox{\delta}) \le \Gamma^{(2,2)}(0),
\end{equation}
where 
\begin{equation}
\label{eq2}
\Gamma^{(2,2)}(\bbox{\rho_1}, \bbox{\rho}_2, \tau) = \langle \bbox{I}(\bbox{\rho_1},\tau)
\bbox{I}(\bbox{\rho_2},t+\tau)\rangle,
\end{equation}
$\bbox{\delta} = \bbox{\rho_1} - \bbox{\rho_2}$ is the distance between the two points where the correlation function
is evaluated and $\tau = 0$.

\begin{figure}[h]
%\vspace*{3.5cm}
%\hspace*{.5cm}
%\special{wmf:fig1.wmf x=7cm y=3cm}
\includegraphics*[width=7cm]{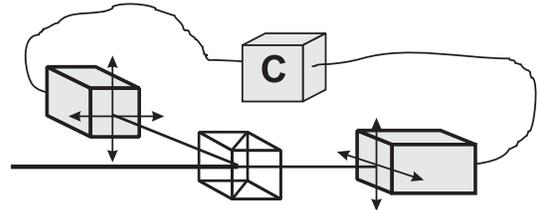}
\caption{Scheme to measure the second order correlation function.}
\label{fig1}
\end{figure}

This kind of correlation function is usually measured with coincidence
photon detection schemes, as described in Fig.\ref{fig1}, for example. The light beam
is sent through a 50/50 beam splitter and coincidence measurements are performed
between detections at detectors D1 and D2 for different positions of these detectors
in the transverse plane. Therefore, the inequality in Eq.\ref{eq1} would be violated
whenever the coincidence counting rate for displaced (in the transverse detection plane)
detectors is bigger than that for aligned ones. 

According to previous works\cite{8}, it is possible to manipulate the transverse profile
of the coincidence counting rate between signal and idler photons from the parametric
down-conversion, by manipulating the pump beam profile. Therefore, we would be able
to prepare collinear twin beams that would give rise to a transverse coincidence
distribution which would violate the inequality in Eq.\ref{eq1}. This requirement
is easily fulfilled preparing the pump beam with zero intensity in its center.
As this intensity distribution is transferred to the coincidence distribution, we would
have zero coincidences for aligned detectors and some higher coincidence counting rates
for displaced detectors.

There is, however, one problem. The violation of the inequality only results in the
spatial photon antibunching if the light beam is homogeneous. This means
that the intensity (second order correlation function) should be constant along the detection
region and the coincidence counting rate (fourth order correlation function) must depend only 
on $\delta$, the relative displacement between detectors D1 and D2. This last requirement
is not fulfilled by the state preparation by manipulation of the pump beam. The coincidence
counting rate depends on the pump beam profile in the sum of the spatial variables\cite{8}:

\begin{equation}
\label{eq3}
C(\rho_i,\rho_s) \propto \left|{\cal W} (\frac{\rho_{s}}{\mu_s}+\frac{\rho_{i}}{\mu_i})\right|^2,
\end{equation}
where ${\cal W}$ is the field transverse distribution of the pump, $\rho_i$ and $\rho_s$
are the transverse coordinates at the detection plane, and $\mu_i$ and $\mu_s$ are
coefficients depending on the distances between signal and idler detection planes to the
crystal.

The dependence of the coincidence counting rate on the sum of the position variables
above, characterizes the non-homogeneity of this field. Therefore, the violation of the
inequality in Eq.\ref{eq1}, does not imply in spatial antibunching. In the following, we
propose and implement a simple method for obtaining homogeneity and as a result, the
antibunching.

\begin{figure}[h]
%\vspace*{3cm}
%\special{wmf:fig2.wmf x=8.5cm y=3cm}
\includegraphics*[width=8cm]{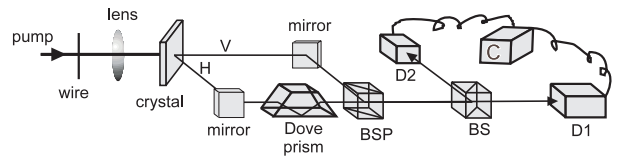}
\caption{Experimental setup.}
\label{fig2}
\end{figure}

The experimental set-up is shown in Fig.\ref{fig2}. Before the pump laser reaches the non-linear crystal, it is passed through a thin wire and through an imaging lens, so that after pumping the crystal, the image of the wire is formed in a plane situated at a certain distance from it. This distance is the same as the distance between crystal and detectors. It has been demonstrated that the coincidence counting rate between signal and idler photons, has a transverse distribution that mimics the pumping beam intensity distribution\cite{8}. This will lead to a coincidence counting rate which is zero when the detectors are aligned and increase when they are displaced. As stated above, this field is non-homogeneous.

In order to overcome this difficulty, we have utilized the scheme displayed in Fig.\ref{fig2}. Signal and idler beams from type II phase matching down-conversion are produced in a non-collinear configuration. The signal beam is passed though a Dove prism, so that its spatial coordinate y is changed into -y. 
As signal and idler have orthogonal polarizations, signal and idler beams can be fully recombined in a polarizing beam splitter. The recombined beam, is split in a non-polarizing beam splitter and the two output beams are sent to detectors. Now, the coincidence counting rate for scans in the y direction is given by:

\begin{equation}
\label{eq4}
C(\rho_i,\rho_s) \propto \left|{\cal W} (\frac{\rho_{s}}{\mu_s}-\frac{\rho_{i}}{\mu_i})\right|^2.
\end{equation}
This characterizes an homogeneous field and as it still keeps the pump image information, it will give rise to spatially anti-bunched photons. The experimental demonstration is easier in one dimension, but the result can be easily extended for the two-dimensional case by insertion of a second Dove prism rotated by 90 degrees with respect to the propagation axis in anyone of the beams.

\section{Experiment}

The experiment have been performed with a cw horizontally polarized HeCd laser operating at 442nm pumping a 1cm long BBO crystal cut for type-II phase matching, as shown in Fig.\ref{fig2}. The down-converted signal and idler beams, at 884nm emerge from the crystal at an angle of 11$^o$ with respect to the laser beam. Two mirrors and a polarizing beam splitter recombine the twin beams. In the signal path a Dove prism is inserted to produce the homogenous field distribution in second order. The combined beam propagates to a non-polarizing beam splitter and are sent to detectors D1 and D2 placed about 75cm from the crystal. Each detector assembly includes a slit of about 0.3mm width, an interference filter centered in 884nm with 10nm bandwidth, a 12.5mm focal length lens, and a single photon counting module.
The detectors are mounted on translation stages. Single and coincidences counts are recorded scanning the detector in y direction.

\section{Results and discussion}

\begin{figure}[h]
%\vspace*{5.5cm}
%\special{wmf:fig3.wmf x=8.5cm y=6cm}
\includegraphics*[width=8cm]{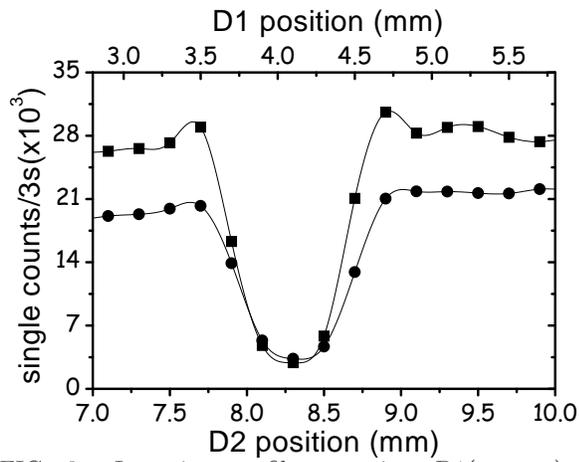}
\caption{Intensity profile scanning D1(squares) and D2(circles).}
\label{fig3}
\end{figure}

\begin{figure}[h]
%\vspace*{12.5cm}
%\special{wmf:fig4.wmf x=8cm y=12cm}
\includegraphics*[width=8cm]{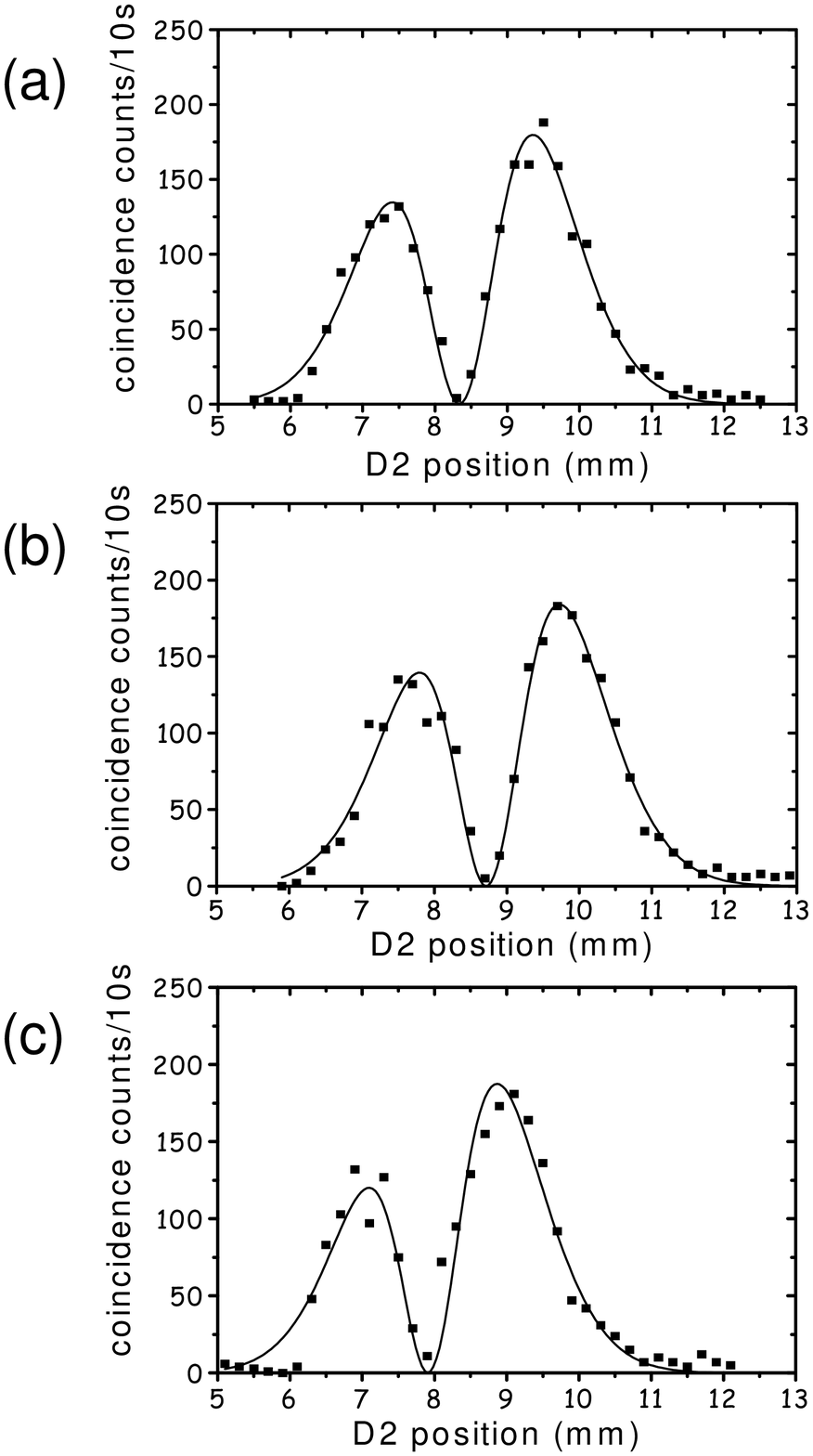}
\caption{Coincidence profile scanning D2 with D1 fixed. (a) D1 at minimum of Fig. 2; (b) D1 displaced +0.4mm; (c) D1 displaced -0.4mm}
\label{fig4}
\end{figure}

\begin{figure}[h]
%\vspace*{12.5cm}
%\special{wmf:fig5.wmf x=8cm y=12cm}
\includegraphics*[width=8cm]{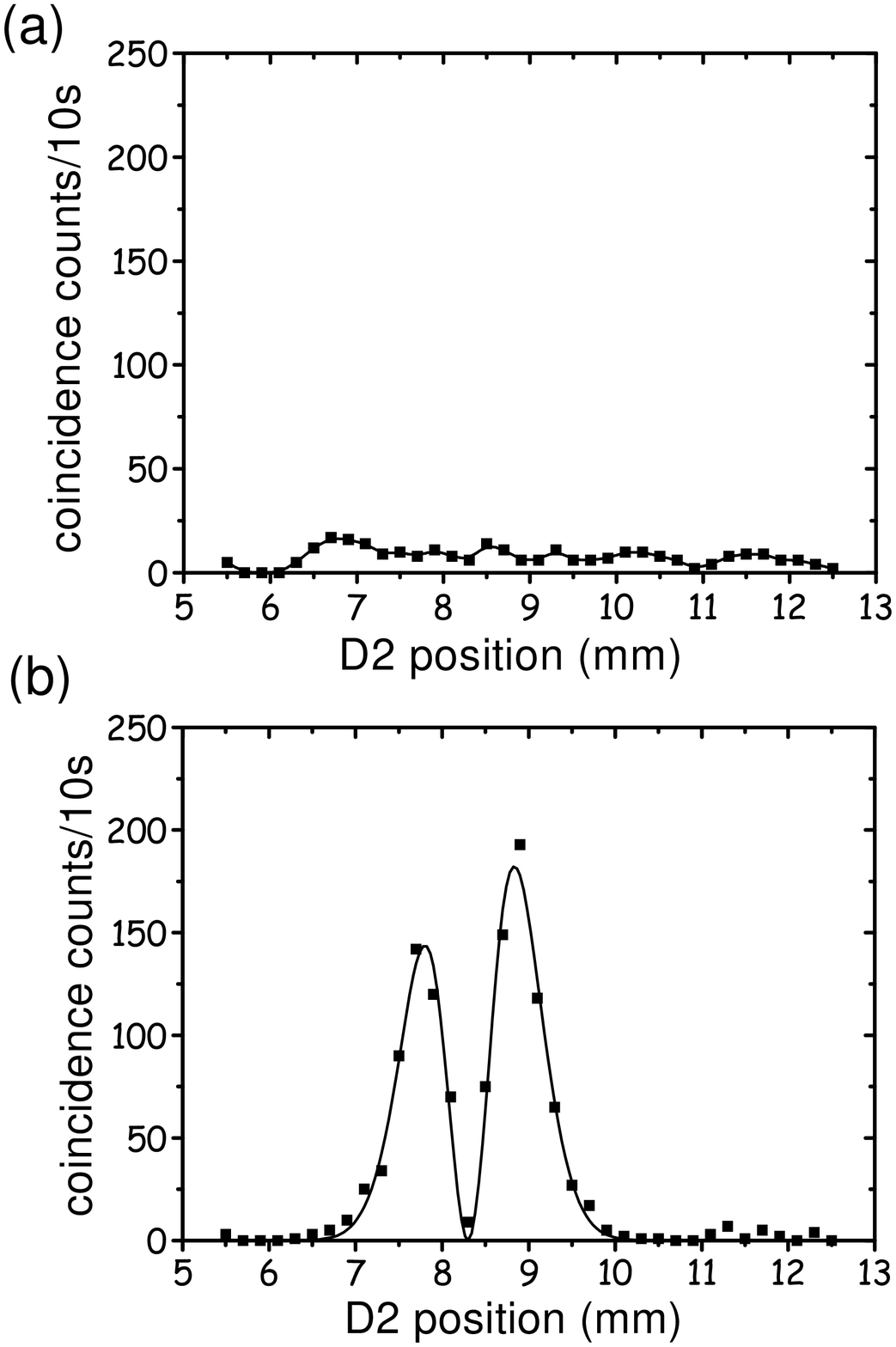}
\caption{Coincidence profile scanning D2 and D1 simultaneously plotted in function of D2 position. (a)same sense; (b) opposite sense.}
\label{fig5}
\end{figure}

We begin by calibrating the position where detectors receive light from the same point of the nonpolarizing beam splitter (BS). This is archived by inserting a thin(500$\mu$m diameter) wire in the beam, before the beam splitter, scanning both detectors, and measuring the single photon counting rates. Fig.\ref{fig3} shows the single counts as a function of D1 and D2 position. In the position where each intensity is a minimum, detectors D1 and D2 will be looking at the same point in the beam splitter, just like a two-photon detector. Now, we remove the wire before the beam-splitter (BS) and place a 250$\mu$m diameter wire and a 25cm focal length lens in the path
of the laser beam before the crystal, such that after the crystal, the image of the wire is formed in a plane situated at the same distance from the crystal as the detectors. In Fig.\ref{fig4}a it is shown the coincidence profile scanning D2 and keeping D1 at the calibration point (minimum of Fig.\ref{fig3}). As we can see, the image of the wire in the laser beam is transferred to the coincidence counting rate, according to the transfer of the angular spectrum\cite{8}. 
We repeat this measurement with D1 displaced +0.4mm and -0.4mm from its previous
position, and the results are shown in Figs.\ref{fig4}b and \ref{fig4}c, respectively. The effect of displacing D1 is the shifting of the coincidence image by the same quantity, showing conditional behavior. Scanning D1 and D2 simultaneously in the same sense, the coincidence counting rate is constant and equal to the minimum of the profile in Fig.\ref{fig4}a, while scanning simultaneously and in opposite senses, the coincidence counting rate depends on the sum of the tranverse coordinates of the signal and idler detectors.
The coincidence profiles for these two situations are shown in Fig.\ref{fig5}. In Fig.\ref{fig5}a we can see a background at the level of the minimum in Fig.\ref{fig4}a, illustrating the homogeneous (dependence only on the difference of the spatial coordinates) character of the field to the forth order, while in the Fig.\ref{fig5}b the image in coincidence appears about two times smaller than that of Fig.\ref{fig4}a.

\begin{figure}[h]
%\vspace*{12.5cm}
%\special{wmf:fig6.wmf x=8cm y=12cm}
\includegraphics*[width=8cm]{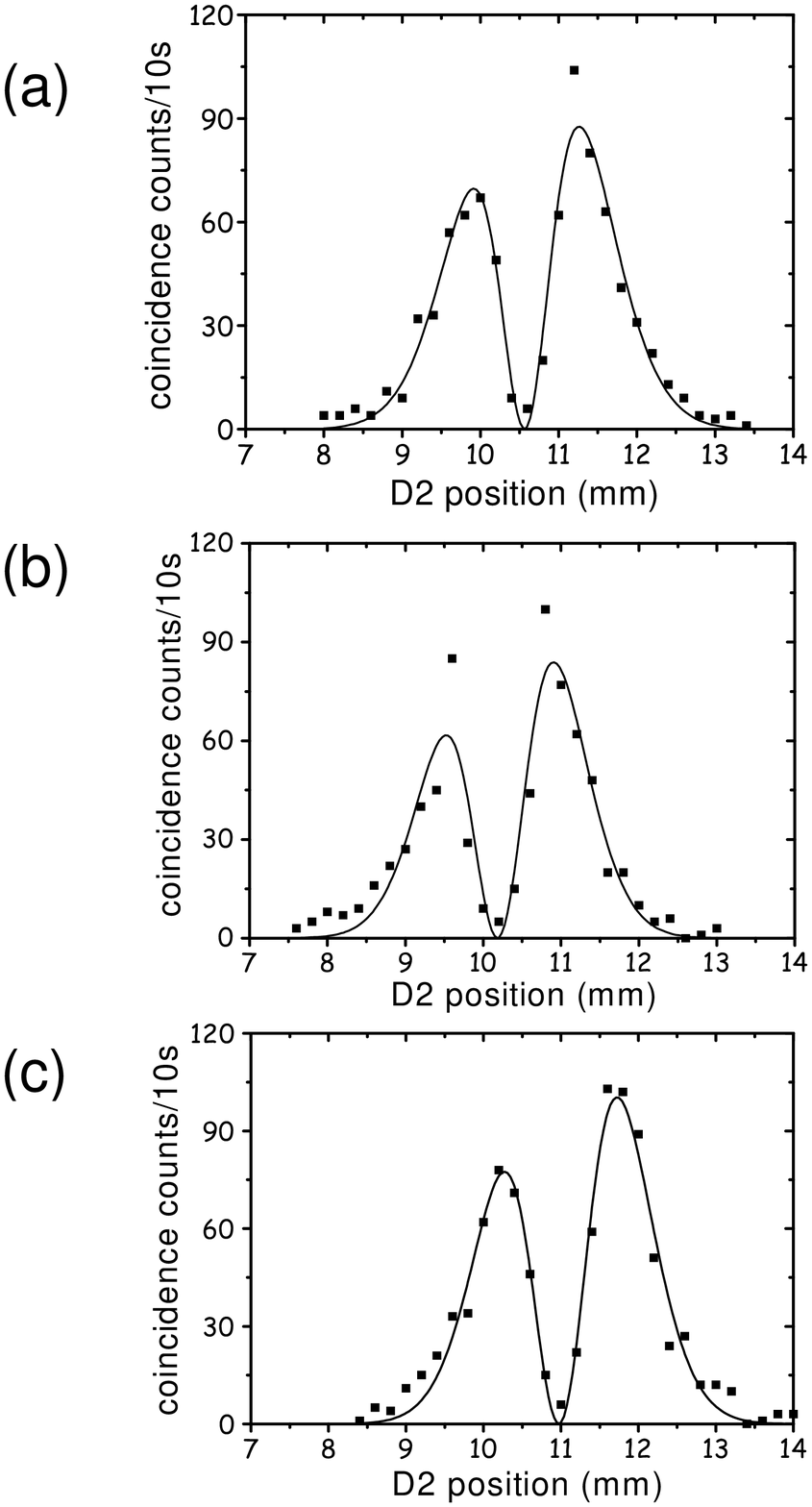}
\caption{Coincidence profile scanning D2 with D1 fixed. (a) D1 at minimum of Fig. 2; (b) D1 displaced +0.4mm; (c) D1 displaced -0.4mm.}
\label{fig6}
\end{figure}

\begin{figure}[h]
%\vspace*{12.5cm}
%\special{wmf:fig7.wmf x=8cm y=12cm}
\includegraphics*[width=8cm]{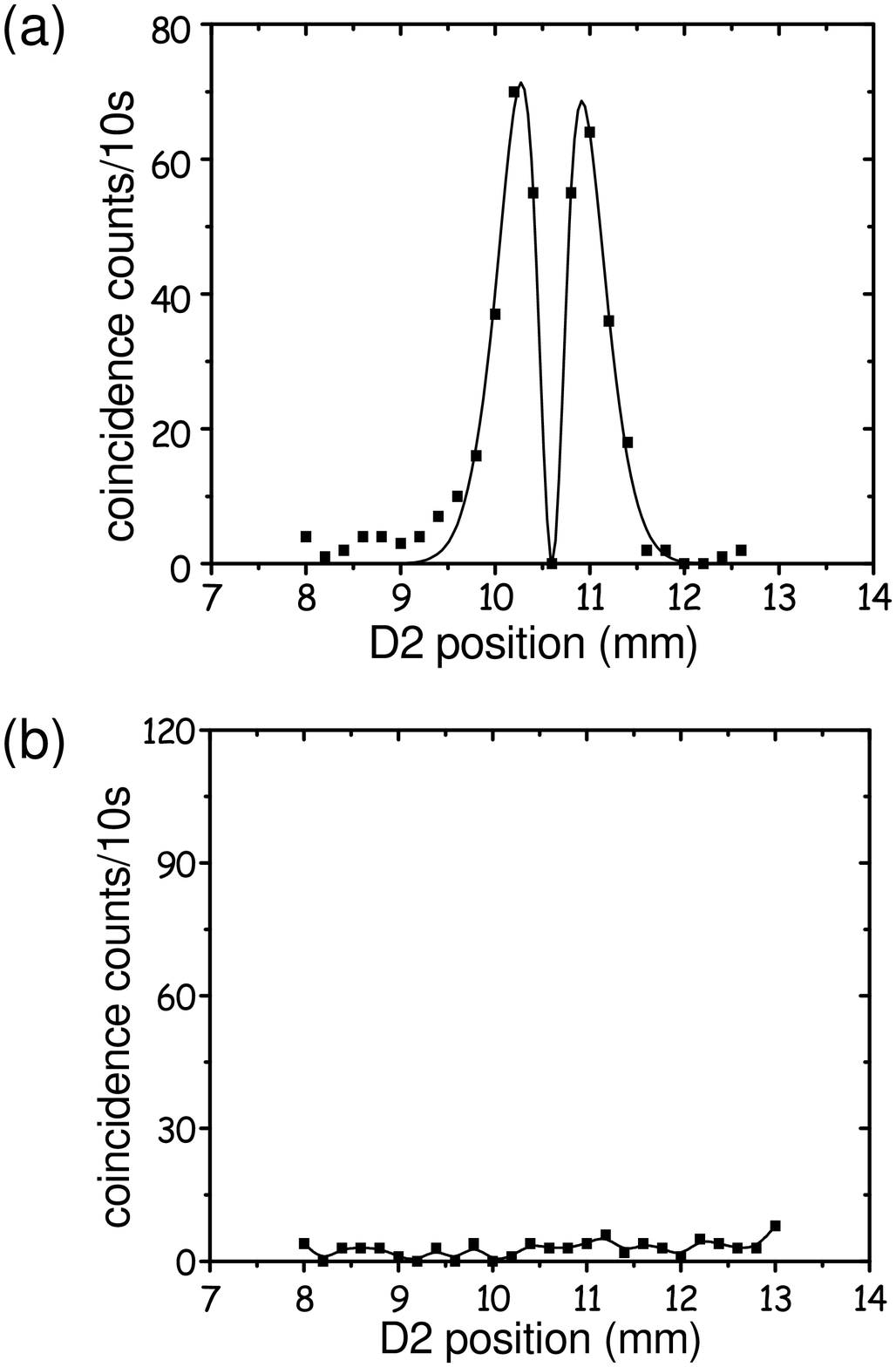}
\caption{Coincidence profile scanning D2 and D1 simultaneously plotted in function of D2 position. (a)same sense; (b) opposite sense.}
\label{fig7}
\end{figure}

We have also  performed the same kind of measurements as before, without the Dove prism. 
The transferred image to the coincidence counting rate is shown in Fig.\ref{fig6}a, while the effect
of the displacement of D1 +0.4mm and -0.4mm from its central position is shown in Figs.\ref{fig6}b 
and \ref{fig6}c, respectively. 
In contrast with the situation described above, the image in coincidence is shifted in
opposite senses with respect to the displacement of D1. 
Scanning D1 and D2 simultaneously, in the same sense, the coincidence counting rate is not 
anymore constant, while scanning in opposite senses it is constant and equal to the minimum of 
Fig.\ref{fig6}a. This results are shown in Fig.\ref{fig7}. 
Here, the coincidence counting rate depends on the sum of the transverse coordinates of the signal and idler beams, so that the field is no longer homogeneous to the fourth order. 
See Fig.\ref{fig7}a. In this case the results cannot be interpreted as spatial antibunching.

One important issue is the fact that no losses are imposed to signal and idler modes during
the preparation process. This improvement might be useful for applications of this kind of
non-classical light, because the coincidence counting rates can be higher than those of
the previous schemes. Another issue is the quality of the state prepared. One way of evaluating 
this quality is through the ratio between the maximum
number of displaced coincident photons(peak of the curves in Figs.\ref{fig4}, for example)  
and the residual number of aligned coincident photons (dip of the curves in Figs.\ref{fig4}, for example). 
This gives us some kind of contrast for the antibunching and our results present high contrasts. 
We are not going to define such a parameter here, because it is not necessary for the purposes of 
this work and that could be confused with the usual contrast defined for interference fringes.

\section{Conclusion}

We have proposed and implemented a scheme for obtaining spatial antibunching
from free-propagating signal and idler beams from the down-conversion. The scheme
consists in inverting the transverse spatial coordinate of one of the twin beams. In 
addition to the preparation of the fourth order correlation between signal and idler
by manipulation of the pump beam, we have succeeded in obtaining an homogeneous
beam that violates a classical inequality and therefore presents spatial antibunching.
This method is very efficient with respect to the initial (just after emission) signal and idler 
beams and presents high quality antibunching.

\begin{acknowledgments}

Financial support was provided by Brazilian agencies CNPq, PRONEX, FAPERJ, FUJB and "Institutos do Mil\^enio" for Quantum Information.

\end{acknowledgments}

\end{document}